\documentclass[pre,twocolumn,amsmath,amssymb,shownopacs]{revtex4}
\usepackage{fancyhdr}
\usepackage{graphicx}
\usepackage{amsmath}
\usepackage{bm}
\usepackage{subfigure}
\usepackage{lastpage}

\begin{document}

\title{Dynamics of a two-level atom in a broadband light field}

\author{Tom \surname{Savels}}
\email{t.savels@utwente.nl}
\author{Allard P. \surname{Mosk}}
\author{Ad \surname{Lagendijk}}
\affiliation{\small{Complex Photonic Systems, Dept. Science and
Technology\\\small University of Twente, PO Box 217, 7500 AE
Enschede, The Netherlands.}}

\date{January 19th, 2004}

\begin{abstract}
We provide a fully analytical description of a two-level atom
interacting with a broadband light field. The problem we present
is a typical example of a physical situation which occurs very
commonly in practice, but is not straightforwardly solvable.
However, we show in this paper that, remarkably, only a limited
number of elementary calculations are required to treat the
problem. The presented results are not only highly intuitive, but
also allow us to study the nontrivial nonlinear response of a
two-level atom to incident broadband radiation.
\end{abstract}

\maketitle \lfoot{} \fancyhead[RO]{\footnotesize{Dynamics of a
two-level system in a broadband field}\\Tom Savels, Allard P. Mosk
and Ad Lagendijk} \cfoot{\thepage\ of \pageref{LastPage}}
\pagestyle{fancy}

\section{\label{sec:intro}Introduction}
The interaction of a two-level atom and a monochromatic field has
always been a very popular topic in quantum physics. The
attraction of this system is owe to the relatively modest
mathematical tools needed to describe the problem, combined with a
rich physical behavior, able to accurately predict many
interesting phenomena such as superradiance \cite{L Allen} or the Mollow triplet \cite{Y Yamamoto}.\\
\indent The effect of the incident field on the atom's dynamics is
described by the optical Bloch equations \cite{M Scully, J-L
Basdevant}. This set of equations allows one to obtain expressions
for the average atomic level populations and coherences in the
presence of an incident field. If the incident light is
monochromatic, the optical Bloch equations are easily dealt with.
For broadband incident light, however, solving the optical Bloch
equations is in general a far from trivial but physically very
relevant matter \cite{A Kaplan, V Chaltykyan, B Blind, H
Freedhoff, T Yoon, S Swain}. For example, a common experimental
situation we are considering is an atom or molecule probed by
near-resonant light, but at the same time irradiated by a
broadband
light source, such as a trapping laser.\\
\indent In this paper, we show how to transparently describe the
interaction of a two-level system and a broadband incident wave.
An everyday (and practically important) example of a broadband
field is the field of spontaneous photons emitted by an atom. The
spectrum corresponding to spontaneous emission has a Lorentzian
frequency distribution, and is therefore broadband. Obviously, we
intuitively expect that if the spectrum of the incident field is
far detuned compared to the resonance frequency of the atom, or
very broad compared to the atom's natural linewidth, no
significant interaction will take place. Likewise, if the incident
field has a very narrow frequency distribution, we expect that the
interaction will resemble the one induced by a monochromatic
incident wave. In this paper, we show that only a few elementary
calculations are required to quantify the interaction of such a
broadband field and a two-level atom. We find that not just the
incident spectrum itself, but in addition the \textit{overlap} of
the incident spectrum and the natural Lorentzian emission line of
the atom itself determines the interaction strength, confirming
our intuitive predictions.\\
\indent In order to simplify the calculations, we will impose a
very general restriction on the incident field. More precisely, we
consider in this paper the class of statistically stationary
incident fields. The mathematical simplifications they allow for,
and the fact that many fields encountered in practice are
statistically stationary, is the reason why they are treated in so
many textbooks on, e.g., quantum optics or magnetic resonance. An
important property of such fields is that two-time averages
$\left<E(t)E(t')\right>$ of the field $E(t)$ only depend on the
time difference $t-t'$, which implies that field components at
different frequencies are
uncorrelated \cite{A Lagendijk}.\\
\indent Finally, we will apply a standard approximation to the
evolution of the atom itself: we will average out the contribution
of highly non-resonating terms. For a monochromatic incident wave,
for example, this approximation simply implies that all components
oscillating at twice the incident wave frequency are neglected.
The physical meaning of this approximation is that non-energy
conserving emission and absorption of photons is neglected, or in
other words, that ``virtual photons'' are not taken into account.
For a monochromatic incident wave, this approximation is referred
to in the literature as the Rotating Wave Approximation (RWA).
Since we deal in this paper with fields which are in general
broadband, we will extend the standard approximation found in
textbooks, and refer to the extension
as the Generalized Rotating Wave Approximation (GRWA).\\
\indent In the following, we will start by formulating the optical
Bloch equations for our configuration. A Fourier transform will
lead to a better understanding of the behavior of our system in
frequency space. Application of the GRWA will return steady-state
solutions for the optical Bloch equations.\\
\indent To conclude our paper, we will show that the results
obtained can be straightforwardly used to express the nonlinear
response of an atom to a broadband field. This response, which can
be quantified by the atom's dynamic polarizability, is of primary
importance in studying atom-field interactions, since it allows
one to gain full insight in the internal atomic dynamics.
\section{\label{sec:Bloch}The optical Bloch equations}
We consider a two-level atom \textit{A} with lower level $a$ and
upper level $c$, separated by an energy difference
$\hbar\omega_{ca}$. The atom interacts with an incident
time-dependent real-valued field ${\bm E}(t)$. The dynamics of the
system can be described in terms of the density matrix
$\hat{\sigma}(t)$. The time evolution of $\hat{\sigma}(t)$ is
given by the optical Bloch equations, which can be written, in the
electric dipole approximation, as
\begin{subequations}
\label{Master Direct, expanded}
\begin{align}
\overset{.}{\sigma}_{cc}&=+i\Omega(t)\Bigl(\sigma_{ca}-\sigma_{ac}\Bigr)-\Gamma\sigma_{cc},\label{Master Direct, expanded,cc}\\
\overset{.}{\sigma}_{aa}&=-i\Omega(t)\Bigl(\sigma_{ca}-\sigma_{ac}\Bigr)+\Gamma\sigma_{cc},\label{Master Direct, expanded,aa}\\
\overset{.}{\sigma}_{ac}&=-i\Omega(t)\Bigl(\sigma_{cc}-\sigma_{aa}\Bigr)+i\omega_{ca}\sigma_{ac}-\frac{\Gamma}{2}\sigma_{ac},\label{Master Direct, expanded,ac}\\
\overset{.}{\sigma}_{ca}&=+i\Omega(t)\Bigl(\sigma_{cc}-\sigma_{aa}\Bigr)-i\omega_{ca}\sigma_{ca}-\frac{\Gamma}{2}\sigma_{ca},\label{Master
Direct, expanded,ca}
\end{align}
\end{subequations}
\\
where the notation $\overset{.}{f}(t)\equiv\frac{d}{dt}f(t)$ is
used. Equations (\ref{Master Direct, expanded}) can be found in
many elementary books on quantum optics \cite{C Cohen-Tannoudji}.
The matrix elements $\sigma_{aa}$ and $\sigma_{cc}$ are the
ensemble-averaged populations of the lower and upper atomic level,
respectively. They are related by $\sigma_{aa}+\sigma_{cc}=1$,
expressing conservation of population. The off-diagonal elements
represent coherences; we will elaborate on their physical meaning
in section \ref{sec:Application}. The constant decay rate $\Gamma$
represents spontaneous emission by the system to the surrounding
vacuum. The Rabi frequency $\Omega(t)\equiv-\frac{1}{\hbar}{\bm
d}_{ac}\cdot{\bm E}(t)$ quantifies the interaction strength
between the atom and the incident field, with ${\bm d}_{ac}$ the
$c \rightarrow a $ transition dipole moment. In the conventional
case of a monochromatic incident wave ${\bm E}(t)\equiv{\bm
E}_{0}\cos\omega_{L}t$ which is often found in the literature, the
definition $\Omega_{Rabi}\equiv-\frac{1}{\hbar}{\bm
d}_{ac}\cdot{\bm E}_{0}$ is mostly used, explicitly removing the
oscillatory time dependence of the field from the Rabi frequency
(obviously, the optical Bloch equations then contain extra factors
$e^{\pm i\omega_{L}t}$). However, we will see that for a more
general time dependence, as the one we deal with here, it is
beneficial to use our definition and consider a time-dependent
Rabi frequency.
\\
\indent Our aim in this section of the paper is to derive
steady-state solutions for equations (\ref{Master Direct,
expanded}). The statistical properties of the field are especially
advantageous in the frequency domain, since if the two-time
average $\left<\Omega(t)\Omega(t+\tau)\right>$ of the phase of
$\Omega$ only depends on $\tau$, one can deduce (see, e.g.,
\cite{A Lagendijk}) that in the frequency domain
\\
\begin{align}
\left<\Omega[\omega]\Omega[\omega']\right>=J[\omega]\delta(\omega+\omega')\label{RPA},
\end{align}
\\
where $J[\omega]$ is the spectral density function of the incident
radiation and the Fourier transform of $\Omega(t)$ is defined as
\\
\begin{subequations}
\begin{align}
\Omega[\omega]&\equiv\frac{1}{2\pi}\int_{-\infty}^{+\infty}\Omega(t)e^{-i\omega
t}dt,\\
\Omega(t)&=\int_{-\infty}^{+\infty}\Omega[\omega]e^{i\omega
t}d\omega.
\end{align}
\end{subequations}
\\
We will see that the appearance of a delta function in (\ref{RPA})
will highly facilitate the calculations further on. In what
follows, we will Fourier transform the optical Bloch equations. We
therefore define the population difference $w(t)$ and its Fourier
transform $w[\omega]$ as
\\
\begin{align}
w(t)&\equiv\sigma_{cc}(t)-\sigma_{aa}(t)\equiv\int_{-\infty}^{+\infty}w[\omega]e^{-i\omega
t}dt\label{Fourier,w},
\end{align}
\\
and
\\
\begin{subequations}
\label{Fourier,coh}
\begin{align}
\sigma_{ac}(t)&\equiv\int_{-\infty}^{+\infty}\sigma_{ac}[\omega]e^{i\omega
t}d\omega\approx\int_{0}^{+\infty}\sigma_{ac}[\omega]e^{i\omega
t}d\omega,\label{Fourier,ac}\\
\sigma_{ca}(t)&\equiv\int_{-\infty}^{+\infty}\sigma_{ca}[\omega]e^{i\omega
t}d\omega\approx\int_{-\infty}^{0}\sigma_{ca}[\omega]e^{i\omega
t}d\omega\label{Fourier,ca},
\end{align}
\end{subequations}
\\
where the contribution of the non-resonating part of the
coherences has been omitted. The restriction of the integration
interval in expressions (\ref{Fourier,coh}) has the simple
physical meaning that non-energy conserving terms are not taken
into account, as mentioned in the introduction of our paper. This
approximation is a straightforward generalization of the Rotating
Wave Approximation \cite{C Cohen-Tannoudji} and is therefore
referred to as the Generalized Rotating Wave Approximation. This
generalization holds excellently if $J[\omega]$ is only
appreciably different from zero near the atomic resonance.\\
\indent If we now split the Fourier transform $\Omega[\omega]$ of
the Rabi frequency $\Omega(t)$ into a positive- and a
negative-frequency part
\\
\begin{align}
\Omega(t)&=\int_{0}^{+\infty}\Omega[\omega]e^{i\omega
t}d\omega+\int_{-\infty}^{0}\Omega[\omega]e^{i\omega
t}d\omega\nonumber\\
&\equiv\Omega_{+}(t)+\Omega_{-}(t),\label{Omega,subst}
\end{align}
\\
we see that neglecting all highly non-resonant terms in
(\ref{Master Direct, expanded}) results in slightly altered
optical Bloch equations:
\\
\begin{subequations}
\label{Master Direct RWA}
\begin{align}
\overset{.}{\sigma}_{cc}&=+i\Omega_{+}(t)\sigma_{ca}-i\Omega_{-}(t)\sigma_{ac}-\Gamma\sigma_{cc},\label{Master Direct RWA, expanded,cc}\\
\overset{.}{\sigma}_{aa}&=-i\Omega_{+}(t)\sigma_{ca}+i\Omega_{-}(t)\sigma_{ac}+\Gamma\sigma_{cc},\label{Master Direct RWA, expanded,aa}\\
\overset{.}{\sigma}_{ac}&=-i\Omega_{+}(t)\Bigl(\sigma_{cc}-\sigma_{aa}\Bigr)+i\omega_{ca}\sigma_{ac}-\frac{\Gamma}{2}\sigma_{ac},\label{Master Direct RWA, expanded,ac}\\
\overset{.}{\sigma}_{ca}&=+i\Omega_{-}(t)\Bigl(\sigma_{cc}-\sigma_{aa}\Bigr)-i\omega_{ca}\sigma_{ca}-\frac{\Gamma}{2}\sigma_{ca}.\label{Master
Direct RWA, expanded,ca}
\end{align}
\end{subequations}
\\
It is clear that equations (\ref{Master Direct RWA}) are far
simpler to deal with than the original Bloch equations
(\ref{Master Direct, expanded}). Fourier transforming equation
(\ref{Master Direct RWA, expanded,cc}) and (\ref{Master Direct
RWA, expanded,aa}) gives
\\
\begin{align}
(\Gamma+i\omega)w[\omega]&=+2i\int_{0}^{+\infty}\Omega[\omega']\sigma_{ca}[\omega-\omega']d\omega'\nonumber\\
&\quad-2i\int_{-\infty}^{0}\Omega[\omega']\sigma_{ac}[\omega-\omega']d\omega'-\Gamma\delta(\omega)\label{w,step1}.
\end{align}
\\
Fourier transforming (\ref{Master Direct RWA, expanded,ac}) and
(\ref{Master Direct RWA, expanded,ca}), on the other hand, gives
\\
\begin{subequations}
\label{Fourier,coh,eqns}
\begin{align}
(i\omega-i\omega_{ca}+\frac{\Gamma}{2})\sigma_{ac}[\omega]&=-i\int_{0}^{+\infty}\Omega[\omega'']w[\omega-\omega'']d\omega''\label{ac,Fourier},\\
(i\omega+i\omega_{ca}+\frac{\Gamma}{2})\sigma_{ca}[\omega]&=+i\int_{-\infty}^{0}\Omega[\omega'']w[\omega-\omega'']d\omega''\label{ca,Fourier}.
\end{align}
\end{subequations}
\\
If we now substitute the previous expressions in (\ref{w,step1}),
we find
\\
\begin{align}
&(\Gamma+i\omega)w[\omega]+\Gamma\delta(\omega)=\nonumber\\\nonumber\\
&=-2\int_{0}^{+\infty}d\omega'\int_{-\infty}^{0}d\omega''\Omega[\omega']\Omega[\omega'']w[\omega-\omega'-\omega'']\times\nonumber\\
&\qquad\Bigl(\frac{1}{i\omega-i\omega'+i\omega_{ca}+\frac{\Gamma}{2}}+\frac{1}{i\omega-i\omega''-i\omega_{ca}+\frac{\Gamma}{2}}\Bigr)\label{w,step2},
\end{align}
\\
which is a self-consistent equation in the population difference.
We use the following trial solution for $w[\omega]$:
\\
\begin{align}
w[\omega]=w[0]\delta(\omega)\label{w,trial},
\end{align}
\\
which is appropriate since we are interested in the regime for
which the populations are time-independent (which is also referred
to as the ``steady-state'' regime). Substitution of
(\ref{w,trial}) in (\ref{w,step2}) and phase-averaging yields
\\
\begin{align}
\Gamma
w[0]+\Gamma&=-2w[0]\int_{0}^{+\infty}d\omega'J[\omega']\frac{\Gamma}{(\omega'-\omega_{ca})^2+(\frac{\Gamma}{2})^2},
\end{align}
\\
and therefore
\\
\begin{align}
w[\omega]&=-\frac{1}{1+2\int_{0}^{+\infty}d\omega'J[\omega']\frac{1}{(\omega'-\omega_{ca})^2+(\frac{\Gamma}{2})^2}}\delta(\omega)\label{w,final}.
\end{align}
\\
We can conclude that in steady-state, we find the desired solution
\\
\begin{subequations}
\label{Bloch_Solutions,final}
\begin{align}
\sigma_{cc}^{st}&\equiv1-\sigma_{aa}^{st}=\frac{X}{2X+1},\\
\sigma_{ca}^{st}&\equiv(\sigma_{ac}^{st})^*=\frac{1}{2X+1}\int_{-\infty}^{0}\frac{i\Omega[\omega]}{-i\omega_{ca}-i\omega-\frac{\Gamma}{2}}e^{i\omega
t}d\omega,
\end{align}
\end{subequations}
\\
with
\\
\begin{align}
X\equiv \int_{0}^{+\infty}d\omega
J[\omega]\frac{1}{(\omega_{ca}-\omega)^2+(\frac{\Gamma}{2})^2}.\label{X}
\end{align}
\\
Expressions (\ref{Bloch_Solutions,final}) are the key result of
our paper. The influence of the spectral properties of the
incident field enters the dynamics of the density matrix through a
single interaction parameter $X$. The expression (\ref{X}) for $X$
is surprisingly simple and appealing: it is the overlap integral
of the spectral density of the incident field, and the natural
Lorentzian emission line of the two-level system itself. The
structure of the interaction parameter confirms what we
intuitively expect: resonant fields with a narrow distribution
interact strongly with the atom, while the interaction with broad,
far off-resonance fields is far less pronounced. As examples, we
will in the next section focus on two specific and interesting
values for $X$.
\section{\label{sec:Examples}Examples}
As a first demonstration of equations
(\ref{Bloch_Solutions,final}), we will verify that the well-known
expressions for an incident \textit{monochromatic} field \cite{C
Cohen-Tannoudji} can be retrieved from our results. We consider a
real-valued monochromatic incident field ${\bm E}(t)\equiv{\bm
E}_{0}\cos(\omega_{L}t)$ with $\omega_L>0$. The corresponding Rabi
frequency is
\\
\begin{align}
\Omega[\omega]\equiv\frac{\Omega_{0}}{2}\Bigl(\delta(\omega-\omega_L)+\delta(\omega+\omega_L)\Bigr),
\end{align}
\\
therefore
\\
\begin{align}
&\Omega[\omega]\Omega[-\omega']=\nonumber\\
&\frac{\Omega_{0}^2}{4}\Bigl(\delta(\omega-\omega')\delta(\omega-\omega_{L})+\delta(\omega-\omega')\delta(\omega+\omega_{L})\nonumber\\
&\quad
+\delta(\omega+\omega')\delta(\omega-\omega_{L})+\delta(\omega+\omega')\delta(\omega+\omega_{L})\Bigr)\label{OmegaOmega,mono,Rotating
Wave}.
\end{align}
\\
Of the 4 terms appearing in (\ref{OmegaOmega,mono,Rotating Wave}),
only the first remains in the (G)RWA, yielding
\\
\begin{align}
J[\omega]=\frac{\Omega_{0}^2}{4}\delta(\omega-\omega_{L})\label{J,mono},
\end{align}
\\
which transforms expressions (\ref{Bloch_Solutions,final}) into
\\
\begin{subequations}
\label{literature}
\begin{align}
\sigma_{cc}^{st}&=(\frac{\Omega_{0}}{2})^2\frac{1}{(\omega_{L}-\omega_{ca})^2+(\frac{\Gamma}{2})^2+\frac{\Omega_{0}^2}{2}},\\
\sigma_{ca}^{st}&=\frac{\Omega_{0}}{2}e^{-i\omega_{L}t}\frac{1}{(\omega_{L}-\omega_{ca})+i\frac{\Gamma}{2}+\frac{1}{2}\frac{\Omega_{0}^2}{(\omega_{L}-\omega_{ca})-i\frac{\Gamma}{2}}},
\end{align}
\end{subequations}
\\
which corresponds exactly to the solutions for incident
monochromatic fields found in the literature \cite{C Cohen-Tannoudji}, justifying our method.\\
\indent As a second demonstration of equations
(\ref{Bloch_Solutions,final}), we consider the nontrivial case of
an atom interacting with incident spontaneous emission. The
incident field then has a Lorentzian spectrum with a width
$\Gamma(1+\kappa)$, $\kappa \geq -1$, centered around
$\omega_{L}\gg\Gamma$:
\\
\begin{align}
J[\omega]\equiv\frac{J_{0}}{\pi}\frac{\frac{\Gamma}{2}(1+\kappa)}{(\omega-\omega_{L})^2+(\frac{\Gamma}{2}(1+\kappa))^2}\label{J,spont},
\end{align}
\\
where the factor
\\
\begin{align}
J_{0}&\equiv\int_{-\infty}^{\infty}J[\omega]d\omega=\left<\Omega(t)^2\right>\label{factor
1_2}
\end{align}
\\
is proportional to the total incident field energy. We find as an
explicit solution for $X$:
\\
\begin{align}
X&=\int_{0}^{+\infty}d\omega J[\omega]\frac{1}{(\omega-\omega_{ca})^2+(\frac{\Gamma}{2})^2}\nonumber\\
&\approx\int_{-\infty}^{+\infty}d\omega
\frac{1}{(\omega-\omega_{L})^2+(\frac{\Gamma}{2}(1+\kappa))^2}
\frac{1}{(\omega-\omega_{ca})^2+(\frac{\Gamma}{2})^2}\times\nonumber\\
&\qquad\frac{J_{0}}{\pi}\frac{\Gamma}{2}(1+\kappa)\nonumber\\
&=J_{0}\frac{(2+\kappa)}{(\omega_{L}-\omega_{ca})^2+(\frac{\Gamma_{ca}}{2})^2(2+\kappa)^2},\label{X2}
\end{align}
\\
where the extension of the integral in (\ref{X2}) from $0$ to
$-\infty$ is clearly justified by $\omega_{L}\gg\Gamma$. Equation
(\ref{X2}) clearly shows that the atom-field interaction is weak
for spectrally broad or far-detuned fields, as mentioned earlier.
To conclude this paper, we will in the next section use the result
(\ref{Bloch_Solutions,final}) with (\ref{literature}) and
(\ref{X2}) obtained here to determine the nonlinear response of an
atom to incident broadband light.
\\
\section{\label{sec:Application}Application}
Let us now focus on a practical application of the previous
sections. If one studies the response of an atom to incident
light, one has to determine the strength with which the atom
interacts with the incident light. A key property quantifying this
strength is the atom's dynamic polarizability $\tensor{\alpha}$.
The associated polarization ${\bm P}$ induced by the incident
field ${\bm E}_{0}$ is then given by ${\bm
P}=\varepsilon_{0}\tensor{\alpha}\cdot {\bm E}_{0}$. In the limit
of low-intensity incident fields, the response is linear and the
polarization is then proportional to the incident field. However,
at higher incident intensities, the response is no longer linear
since the atom will start to saturate. The polarizability will
then depend on the incident field and the polarization will no
longer be proportional to the incident field. We will now show how
the results from the previous sections can be straightforwardly
used to quantify the rate at which the atom saturates due to a
incident (broadband) field. In other words, we will show how
expressions (\ref{Bloch_Solutions,final}) allow us to calculate
the \textit{nonlinear} dynamic polarizability of a
two-level atom.\\
\indent We take the general case of an incident field ${\bm E}(t)$
consisting of a monochromatic cosine component with amplitude
${\bm E}_{0}$ and frequency $\omega_{L}$, and a non-monochromatic
component ${\bm E}_{L}(t)$, with a Lorentzian frequency
distribution. The fields ${\bm E}(t)$ thus obtained define a very
general class, containing the special cases of incident
monochromatic light, as well as pure incident spontaneous
emission. The Lorentzian component of the incident field has a
linewidth $\Gamma(1+\kappa)$, $\kappa\geq -1$, centered around
$\omega_{L}\gg\Gamma$:
\\
\begin{align}
{\bm E}[\omega]={\bm
E}_{0}\frac{1}{2}\Bigl(\delta(\omega+\omega_{L})+\delta(\omega-\omega_{L})\Bigr)+{\bm
E}_{L}[\omega].\label{application}
\end{align}
\\
The corresponding Rabi frequencies are
\\
\begin{subequations}
\begin{align}
-\hbar \Omega_{0}&={\bm d}_{ac}\cdot{\bm E}_{0},\\
-\hbar \Omega_{L}[\omega]&={\bm d}_{ac}\cdot{\bm E}_{L}[\omega],
\end{align}
\end{subequations}
\\
obeying
\\
\begin{align}
\left<\Omega_L[\omega]\Omega_L[\omega']\right>=J_{L}[\omega]\delta(\omega+\omega'),
\end{align}
\\
with $J_{L}[\omega]$ the spectral density of the incident
Lorentzian field
\\
\begin{align}
J_{L}[\omega]\equiv\frac{\mathcal{J}(\kappa)}{\pi}\frac{\frac{\Gamma}{2}(1+\kappa)}{(\omega-\omega_{L})^2+(\frac{\Gamma}{2}(1+\kappa))^2}.\label{Lorentz
}
\end{align}
\\
Equations (\ref{application})-(\ref{Lorentz }) fully describe the
incident field. We will now focus on the response of the atom to
this incident field. Since the dynamic polarizability and the
density matrix both describe the response of the atom to incident
light, both quantities should be connected. Indeed, they are
related by \cite{R Laudon}
\\
\begin{align}
\varepsilon_{0}\tensor{\alpha}(\tilde{\omega})\cdot{\bm
E}[\tilde{\omega}]\equiv{\bm
d}_{ac}\sigma_{ca}[-\tilde{\omega}].\quad
\omega>0,\label{alpha,definition}
\end{align}
\\
Our general result (\ref{Bloch_Solutions,final}) now allows us to
deduce that (\ref{alpha,definition}) leads to
\\
\begin{align}
&\varepsilon_{0}\tensor{\alpha}(\omega)\cdot\Bigl(\frac{1}{2}{\bm
E}_{0}\delta(\omega+\omega_{L})+{\bm
E}_{L}[\omega]\Bigr)\nonumber\\
&=-\frac{{\bm
d}_{ac}}{2X+1}\Bigl(\frac{1}{2}\frac{\Omega_{0}\delta(\omega+\omega_{L})}{{\omega_{ca}-\omega-i\frac{\Gamma}{2}}}+\frac{\Omega_{L}[\omega]}{\omega_{ca}-\omega-i\frac{\Gamma}{2}
}\Bigr)\nonumber\\
&=\frac{{\bm d}_{ac}}{2X+1}\cdot\Bigl(\frac{1}{2}\frac{{\bm
d}_{ac}\cdot{\bm
E}_{0}\delta(\omega-\omega_{L})}{{\omega_{ca}-\omega-i\frac{\Gamma}{2}}}+\frac{{\bm
d}_{ac}\cdot{\bm
E}_{L}[\omega]}{\omega_{ca}-\omega-i\frac{\Gamma}{2} }\Bigr),
\end{align}
\\
where $\otimes$ represents the tensor product of two vectors. The
nonlinear polarizability is thus expressed as
\\
\begin{align}
\tensor{\alpha}(\omega)&=\frac{{\bm d}_{ac}\otimes{\bm d}_{ac}}{\varepsilon_{0}\hbar}\frac{1}{2X+1}\frac{1}{\omega_{ca}-\omega-i\frac{\Gamma}{2}\nonumber}\\
&=-\tensor{\alpha}_{0}\frac{1}{2}\frac{\omega_{ca}}{\omega-\omega_{ca}+i\frac{\Gamma}{2}+\frac{2S}{\omega-\omega_{ca}-i\frac{\Gamma}{2}}}\label{T},
\end{align}
\\
where the static polarizability is given by
\\
\begin{align}
\tensor{\alpha}_{0}\equiv\frac{2}{\omega_{ca}\hbar\varepsilon_{0}}{\bm
d}_{ac}\otimes{\bm d}_{ac},
\end{align}
\\
and where the saturation parameter can be written in a
surprisingly simple way as
\\
\begin{align}
S&\equiv \frac{\Omega_{0}^2}{4}+\int_{0}^{+\infty}d\omega
J_{L}[\omega]\frac{{(\omega_{L}-\omega_{ca})^2+(\frac{\Gamma}{2})^2}}{(\omega-\omega_{ca})^2+(\frac{\Gamma}{2})^2}\label{S}.
\end{align}
\\
Expressions (\ref{T})-(\ref{S}) fully describe the response of a
two-level atom to an incident field of the general class
(\ref{application}).\\
\indent Two limits for the incident field are in particular
interesting. Firstly, for $J_{L}[\omega]\rightarrow 0$, the
expression for the dynamic polarizability of a two-level atom
irradiated by a monochromatic field is recovered
\\
\begin{align}
\tensor{\alpha}(\omega)&=-\tensor{\alpha}_{0}\frac{1}{2}\frac{\omega_{ca}}{\omega-\omega_{ca}+i\frac{\Gamma}{2}+\frac{\Omega_{0}^2/2}{\omega-\omega_{ca}-i\frac{\Gamma}{2}}}\label{alpha,mono}.
\end{align}
\\
\indent Secondly, for $\Omega_{0}\rightarrow 0$, we find the
saturation due to pure spontaneous emission
\\
\begin{align}
\tensor{\alpha}(\omega_{L})&=-\tensor{\alpha}_{0}\frac{1}{2(2X+1)}\frac{\omega_{ca}}{\omega_{L}-\omega_{ca}+i\frac{\Gamma}{2}}\label{alpha,spont},
\end{align}
\\
with $X$ given by expression (\ref{X}).\\
\indent Finally, we note that expressions (\ref{alpha,mono}) and
(\ref{alpha,spont}) imply that for small incident fields, the same
expression
\\
\begin{align}
\tensor{\alpha}(\omega_{L})&=-\tensor{\alpha}_{0}\frac{1}{2}\frac{\omega_{ca}}{\omega_{L}-\omega_{ca}+i\frac{\Gamma}{2}}\label{alpha,linear},
\end{align}
\\
for the \textit{linear} dynamic polarizability \cite{P d Vries} is
obtained, as it should be.
\section{\label{sec:Summary}Summary}
In this paper, we have solved the optical Bloch equations for a
two-level system interacting with a statistically stationary
broadband field. The resulting steady-state density matrix is
similar to the result found for a monochromatic incident field;
the difference between both results can be intuitively understood.
Finally, we have applied the obtained results to calculate the
full response of a two-level atom to a broadband field, expressed
by the atom's dynamic polarizability.

\begin{acknowledgements}
This work is part of the research program of the `Stichting voor
Fundamenteel Onderzoek der Materie' (FOM), which is financially
supported by the `Nederlandse Organisatie voor Wetenschappelijk
Onderzoek' (NWO).
\end{acknowledgements}

\end{document}